# Spatial intensity distribution of light under coherent interaction with single atoms

Akifumi Takamizawa[1*] and Koichi Shimoda[2†]

[1]*National Metrology Institute of Japan (NMIJ), National Institute of Advanced Industrial Science and Technology, Tsukuba, Ibaraki 305-8563, Japan*

[2]*The Japan Academy, Ueno Park, Tokyo 110-0007, Japan*

[*] Email: akifumi.takamizawa@aist.go.jp

[†] Professor emeritus at The University of Tokyo

# Spatial intensity distribution of light under coherent interaction with single atoms

Variations in the spatial intensity distribution of light caused by coherent interaction with two-level atoms are determined by semi-classically calculating a term for interference between incident light and spherical radiation from the atom. The interference term is divided into two components: one oscillates at the Rabi frequency corresponding to absorption and stimulated emission, and the other contributes to dispersion. The spatial distribution of both components involves alternation of constructive and destructive parts in the form of parabolic surfaces where light energy flows. Although the time average over the Rabi oscillation of the former component vanishes at every position, that of the latter remains. The light-related reaction of a dipole force acting on an atom is considered to appear in the non-vanishing latter component.

## 1. Introduction

A promising way of developing quantum information technology involves the use of strong interaction between single atoms and light. In this context, the application of a cavity quantum electrodynamics effect with a high-finesse cavity to bring about such interaction has been widely investigated [1, 2]. Strong interaction in free space using tightly focused light has also been studied recently [3 – 10]. To support such studies, a deep understanding of interaction between atoms and light has been necessary. Particularly for strong interaction in free space, both the power and the spatial intensity distribution of light are significant considerations [3 – 6].

In addition to the investigation of single molecules in solids or liquids [11, 12], absorption images of single atomic ions isolated in a vacuum have recently been observed [13]. In this experiment, absorption of cooling light focused on an ion was visualized in image form with near-wavelength resolution. Here, two-dimensional distribution is more informative than data on the power of absorption alone. It should be noted that an absorption image of a single atom is essentially different from one of an atomic ensemble: the former encompasses the electric dipole induced in the atom, while the latter simply shows the density distribution of the ensemble. Observation of the spatial intensity distribution of absorption or stimulated emission is expected to provide new and powerful resources to support investigation of interaction between single atoms and light.

In the present study, variations in the spatial intensity distribution of plane-wave incident light caused by coherent interaction with two-level atoms were semi-classically calculated and discussed. Here, spatial intensity distribution under coherent interaction is expressed by superposition between the electromagnetic field of incident light and that of spherical radiation from the electric dipole induced in the atom (see Appendix 1 for an intuitive outline of the concept). Accordingly, variations in the spatial intensity distribution of incident light are given by the interference term of the superposition. Thus, the phase relationship between incident light and the electric dipole plays an important role, in contrast to the calculation of an emitter's spontaneous decay rate modified by a metallic surface or nanoparticle [14 – 18]. The proposed calculation shows the dependence of spatial intensity distribution on parameters such as time and frequency detuning. In addition to stimulated emission and absorption, dispersion-related variations in spatial intensity distribution are also represented.

Results obtained using the proposed calculation method agreed qualitatively with the absorption image in the experiment described in Ref. [13]. It can also be modified to suit various situations, such as cases where incident light is tightly focused on an atom.

In Section 2, calculation of a term for interference between incident light and spherical radiation from the atom is discussed. Section 3 covers spatial intensity distribution and its time dependence, and includes consideration of how the dipole force acting on an atom reacts to the light field as well as comparison with the experiment described in Ref. [13]. Section 4 outlines the conclusions of the study.

## 2. Interference term calculation

For theoretical calculation, a case is considered in which incident light whose electric field at time $t$ is expressed as $\boldsymbol{E}_L = E_0 \cos(\omega t - kx)\hat{z}$ illuminates a two-level atom in a vacuum, where the origin of inertial Cartesian coordinates ($x$, $y$, $z$) is taken as the position of the atom. Here, $E_0$, $\omega$, $k$ and $\hat{z}$ are the amplitude of the electric field, the frequency, the wave number in the vacuum and the unit vector in the $z$ direction, respectively. The wave function of the two-level atom is expressed as $\psi = a_1(t)\psi_1 + a_2(t)\psi_2$, where $\psi_1$ and $\psi_2$ are the eigenfunction of the ground state and that of the excited state, respectively.

Here, a simple case with only coherent interaction is considered, neglecting spontaneous emission. As a result of coherent interaction due to the perturbation Hamiltonian $H' = -\mu_z E_L$ (where $\mu_z$ is the $z$ component of the electric dipole moment) under the initial conditions of $a_1(0) = 1$ and $a_2(0) = 0$, the quantum-mechanical expectation value of the induced dipole $\int \psi^* e z \psi d\mathbf{r}$ (where $e$ is the electron charge) is given by $\boldsymbol{p}(t) = p_0(t)\exp(i\omega t)\hat{z} + c.c.$ [19], where

$$p_0(t) = -\frac{\Omega_0}{2\Omega}|\mu_{21}|\left[\frac{\Delta}{\Omega}(1-\cos\Omega t) + i\sin\Omega t\right]. \tag{1}$$

Here, $\mu_{21}$ and $\Delta = \omega - \omega_0 + \delta_D$ are the matrix element of $\mu_z$ and the frequency detuning of incident light with respect to the atomic resonant frequency $\omega_0$, respectively, where $\delta_D$ is the Doppler shift caused by the movement of the atom with respect to the light source. In addition, $\Omega_0 = |\mu_{21}E_0|/\hbar$ is the vacuum Rabi frequency, where $2\pi\hbar$ is Planck's constant $h$ and $\Omega = (\Delta^2 + \Omega_0^2)^{1/2}$ is the Rabi frequency.

The light intensity given by superposition between incident light and radiation from the induced dipole is expressed as $|E|^2 = |E_L|^2 + |E_p|^2 + \eta(r,t)$. Here, $\eta(r,t) = 2E_L \cdot E_p$ is the interference term at position $r$, where $E_p$ is the electric field of the spherical radiation from the dipole. This is rewritten as $\eta(r,t) = \eta_r(r,t) + \eta_i(r,t)$, where $\eta_r(r,t)$ and $\eta_i(r,t)$ are the functions of $\text{Re}[p_0(t)]$ and $\text{Im}[p_0(t)]$, respectively. When $\omega \gg \Omega$ and $kR \gg 1$, they are expressed as

$$\eta_r(r,t) = -A\frac{\Delta}{\Omega} \cdot \frac{\sin^2\theta}{kR}\cos(kR-kx)(1-\cos\Omega t), \tag{2a}$$

$$\eta_i(r,t) = -A\frac{\sin^2\theta}{kR}\sin(kR-kx)\sin\Omega t. \tag{2b}$$

Here, $R = (x^2 + y^2 + z^2)^{1/2}$ and $\theta = \cos^{-1}(z/R)$. The constant $A$ is given by $A = \hbar k^3 \Omega_0^2/(4\pi\varepsilon_0\Omega)$, where $\varepsilon_0$ represents permittivity in a vacuum. Assuming that $|E_L| \gg |E_p|$, the interference term gives the variation in light intensity caused by coherent interaction. Here, $|E_p|^2$ is the intensity of Rayleigh scattering [20]. It should be noted that the condition $|E_L| \gg |E_p|$ is usually satisfied where $kR \gg 1$ because the power of Rayleigh scattering is as small as $\sim \hbar\omega\Gamma/8$ (where $\Gamma$ is the natural linewidth in an exited state) [21].

While $\text{Re}[p_0(t)]$ corresponds to dispersion, $\text{Im}[p_0(t)]$ contributes to absorption and stimulated emission. Thus, power enhancement or reduction is expressed not by the component $\eta_r(\boldsymbol{r},t)$ but by $\eta_i(\boldsymbol{r},t)$. As found from Eqs. (2a) and (2b), the components $\eta_r(\boldsymbol{r},t)$ and $\eta_i(\boldsymbol{r},t)$ both oscillate at the Rabi frequency.

## 3. Discussion

### *3.1 Stimulated emission and absorption*

Here, the component $\eta_i(\boldsymbol{r},t)$ is discussed. Figure 1(a) shows $\eta_i(\boldsymbol{r},t)$ for $kz = 0$ and $\Omega t = 3\pi/2$ as functions of $kx$ and $ky$. Figure 1(b) represents $\eta_i(\boldsymbol{r},t)$ for $kx = 50$ and $kz = 0$ as a function of $ky$, where the parameter is $\Omega t$. Figures 1(c) and (d) show $\eta_i(\boldsymbol{r},t)$ for $kx = 50$ as functions of $ky$ and $kz$, where $\Omega t = 3\pi/2$ and $\Omega t = \pi/2$, respectively. It should be noted that stimulated emission and absorption reach their maxima at $\Omega t = 3\pi/2$ and $\Omega t = \pi/2$, respectively, as the power of stimulated emission is given by $\omega E_0 \text{Im}[p_0(t)]$.

As shown in Figs. 1(a) – (d) and as found from Eq. (2b), the spatial distribution of $\eta_i(\boldsymbol{r},t)$ involves alternation of constructive and destructive parts at all times except for $\Omega t = n\pi$ (where $n$ is an integer). Thus, it is found that the spatial intensity distribution of stimulated emission (absorption) is not homogeneous over the whole space.

It can be seen from Fig. 1(a) that constructive and destructive parts appear along curved surfaces. Here, the surface is expressed as $kR - kx = C$, where $C$ is zero or a positive constant. Using $\rho = \left(y^2 + z^2\right)^{1/2}$ gives

$$\begin{aligned} x &= \frac{k}{2C}\rho^2 - \frac{C}{2k} \quad (C > 0) \\ \rho &= 0 \quad\quad\quad\quad\quad\;\; (C = 0) \end{aligned} \qquad (3)$$

When $\sin\Omega t < 0$ ($\sin\Omega t > 0$), the component $\eta_i(\boldsymbol{r},t)$ is the most constructive (destructive) at $C = \pi/2 + 2n\pi$ and the most destructive (constructive) at $C = 3\pi/2 + 2n\pi$, as given by Eq. (2b). Accordingly, Eq. (3) shows that constructive and destructive parts can be expressed by parabolic surfaces that are concentric around the *x* axis (see Figs. 1(a) – (d)). The component $\eta_i(\mathbf{r},t)$ is then zero on the *x* axis at any time in the region for $x > 0$.

As found from Eq. (2b), the time dependence of $\eta_i(\mathbf{r},t)$ is given by $\sin\Omega t$ independently of the position $\mathbf{r}$. Hence, as shown in Fig. 1(b) and as indicated from comparison between Figs. 1(c) and (d), the spatial pattern of $\eta_i(\mathbf{r},t)$ is independent of time except for the sign. At arbitrary positions, the component $\eta_i(\mathbf{r},t)$ oscillates at the Rabi frequency, and the time average of $\eta_i(\mathbf{r},t)$ over the Rabi oscillation, $\overline{\eta_i(\mathbf{r},t)}$, is zero at all positions.

*3.2 Dispersion*

Here, the component $\eta_r(\mathbf{r},t)$ is discussed. It is known from Eq. (2a) that the sign of $\eta_r(\mathbf{r},t)$ changes with the sign of frequency detuning $\Delta$, and that $|\eta_r(\mathbf{r},t)|$ reaches its maximum at $\Omega t = \pi + 2n\pi$. Figures 2(a) and (b) show $\eta_r(\mathbf{r},t)$ for $\Delta > 0$ and $\Delta < 0$, respectively, for $kx = 50$ and $kz = 0$ as functions of $ky$, where the parameter is $\Omega t$. Figures 2(c) and (d) also display $\eta_r(\mathbf{r},t)$ for $\Delta > 0$ and $\Delta < 0$, respectively, for $kx = 50$ and $\Omega t = \pi$ as functions of $ky$ and $kz$.

As with the component $\eta_i(\mathbf{r},t)$, $\eta_r(\mathbf{r},t)$ has alternate constructive and destructive parts in the form of parabolic surfaces, except that one of the destructive (constructive) parts is on the *x* axis in the region for $x > 0$ and $\Delta > 0$ ($\Delta < 0$). However, in contrast to $\eta_i(\mathbf{r},t)$, $\eta_r(\mathbf{r},t)$ oscillates with no change of sign at all positions (as shown in Figs. 2(a) and (b)) because the time dependence is given by $1 - \cos\Omega t$ from Eq. (2a). Accordingly, the time averages of $\eta_r(\mathbf{r},t)$ over the Rabi oscillation, $\overline{\eta_r(\mathbf{r},t)}$, are non-vanishing at all positions except where $\cos(kR - kx) = 0$.

For a case where $|\Delta| \ll \Omega_0$, Figs. 1(a) – (d) show the spatial distribution of the interference term $\eta(\boldsymbol{r},t)$, as $\eta_\mathrm{r}(\boldsymbol{r},t)$ is negligible compared to $\eta_\mathrm{i}(\boldsymbol{r},t)$, as found from Eqs. (2a) and (2b). For a case where $|\Delta| \gtrsim \Omega_0$, on the other hand, the spatial distribution and the time variation of $\eta(\boldsymbol{r},t)$ are rather complicated, as $\eta_\mathrm{r}(\boldsymbol{r},t)$ cannot be ignored. Nevertheless, numerical calculation shows that the spatial distribution of $\eta(\boldsymbol{r},t)$ is concentric around the $x$ axis and involves alternation of constructive and destructive parts (see Fig. 3 for examples).

*3.3 Reaction of dipole force*

As shown by Eqs. (2a) and (2b), the amplitudes of the components $\eta_\mathrm{r}(\boldsymbol{r},t)$ and $\eta_\mathrm{i}(\boldsymbol{r},t)$ both increase simply with $E_0$. Hence, the spatial distribution of $\eta(\boldsymbol{r},t)$ depends on that of the incident light field. However, in terms of the time average over the Rabi oscillation, $\overline{\eta_\mathrm{i}(\boldsymbol{r},t)}$ is zero independently of $E_0$, whereas $\overline{\eta_\mathrm{r}(\boldsymbol{r},t)}$ is a function of $E_0$.

Accordingly, if $E_0$ is not symmetric around the $x$ axis, the spatial distribution of $\overline{\eta_\mathrm{r}(\boldsymbol{r},t)}$ will be unbalanced. This imbalance can be regarded as a light-related reaction of the dipole force acting on an atom. In fact, the dipole force results from $\mathrm{Re}[p_0(t)]$ rather than from $\mathrm{Im}[p_0(t)]$, and is proportional to $\nabla E_0$. In other words, it is considered that the dipole force can be directly measured from the spatial imbalance of $\overline{\eta_\mathrm{r}(\boldsymbol{r},t)}$.

In an optical dipole trap with far-detuned light [22, 23], fluorescence induced by trapping light is ordinarily too weak to be detected. Hence, additional resonant light is shone on atoms to induce fluorescence, which degrades the trap due to spontaneous emission [24]. The measurement of $\overline{\eta_\mathrm{r}(\boldsymbol{r},t)}$ for the trapping light will enable non-destructive detection of atoms. It should be noted that no additional probing light is necessary in contrast to other non-destructive detection methods such as observation of fluorescence excited by cooling light [25, 26] and measurement of the phase shift of additional off-resonant light [27].

*3.4 Comparison with experimental results*

A central dark spot and a surrounding ring-like pattern can be clearly seen in the absorption image shown in Fig. 2b in Ref. [13]. In this experiment, cooling light with a wavelength of $\lambda = 369.5$ nm and an intensity of $I \cong I_s$ (where $I_s$ is the saturation intensity) was focused on and absorbed by a $^{174}$Yb$^+$ ion. An absorption image on the $x = 0$ plane was captured using a phase Fresnel lens, where $\Delta = -8$ MHz $\cong -0.3\Gamma$ [28]. It is then given that $\Omega \cong 0.8\Gamma$ based on $\Omega_0 = [I/(2I_s)]^{1/2}\Gamma$ [29], so the stimulated emission rate was comparable with the spontaneous emission rate. Thus, absorption exceeded stimulated emission, and values of $\overline{\eta_i(r,t)} \sim \eta_i(r, \pi/(2\Omega))$ applied. On the $x = 0$ plane, $\overline{\eta_i(r,t)} \propto -\sin^2\theta \sin(k\rho)/k\rho$ was true for $k\rho \gg 1$. The component $\overline{\eta_r(r,t)}$ was comparable to $\overline{\eta_i(r,t)}$, as $|\Delta/\Omega| \cong 0.4$. The component $\overline{\eta_r(r,t)}$ may therefore have contributed to the absorption image.

Unfortunately, it is difficult to quantitatively discuss the components $\overline{\eta_i(r,t)}$ and $\overline{\eta_r(r,t)}$ with comparison to the experimental results, especially in regard to the central dark spot where $\rho < \lambda$, because calculation on the $x = 0$ plane is performed under a condition in which $k\rho \gg 1$ and the resolution of the image is restricted by the diffraction limit. Nevertheless, qualitative correspondence regarding the characteristic ring-like pattern with a one-wavelength interval is seen. This pattern is incorporated into the proposed calculation via interference between incident light and spherical radiation from the atom. This shows that the phase relationship between incident light and the induced electric dipole is important in consideration of spatial intensity distribution under atom-light interaction.

## 4. Conclusion

In this study, variations in the spatial intensity distribution of light caused by absorption, stimulated emission and dispersion were ascertained by calculating a term for interference between incident light and spherical radiation from the atom. It was confirmed that the spatial distributions of the two interference term components $\eta_r(\boldsymbol{r},t)$ and $\eta_i(\boldsymbol{r},t)$ involved alternation of constructive and destructive parts along parabolic surfaces. A ring-like pattern that appeared on the plane vertical to the wave vector of incident light was found to correspond to the characteristics of the absorption image in the experiment. In terms of time averages over the Rabi frequency, $\overline{\eta_r(\boldsymbol{r},t)}$ does not vanish, while $\overline{\eta_i(\boldsymbol{r},t)}$ does. Taking advantage of spatial intensity distribution, it is considered that non-destructive detection of atoms in an optical dipole trap and direct measurement of dipole force can be achieved.

Acknowledgements

The authors would like to thank Prof. Kielpinski at Griffith University for providing helpful comments on the revision of this manuscript, and are also grateful to Dr. Tanabe at the National Metrology Institute of Japan for helping with the introduction.

**Appendix 1: Intuitive outline of spatial intensity distribution**

Figure 4 shows a schematic pattern of light waves under coherent interaction with an atom [30]. Here, the incident light on resonance to the atom propagating toward the right illuminates the atom located in the center. Here, the straight (black) and circular (blue) lines show the wave fronts of incident light and those of spherical radiation from the atom, respectively. The parabolic (red) lines connecting the intersections of the straight lines with the circular lines represent the surfaces on which the incident light and spherical radiation have the same phase and interfere with each other constructively. The energy of the stimulated emission flows along the parabolic lines and finally travels in the direction of the incident light.

It should be noted that the phase of spherical radiation is advanced with respect to that of incident light by $\pi/2$ in Fig. 4. In this case, as the power of stimulated emission given by $\omega E_0 \text{Im}[p_0(t)]$ is positive, the transition from an excited state to a ground state occurs via stimulated emission. In contrast, when incident light is absorbed by the atom, this light and spherical radiation interfere with each other destructively on the parabolic lines because the phase of the radiation is retarded with respect to that of the light by $\pi/2$. Stimulated emission and absorption are periodically repeated at the Rabi frequency.

**Figure Captions**

Figure 1. Spatial distribution of $\eta_i(\boldsymbol{r},t)$ – component of interference term $\eta(\boldsymbol{r},t)$ contributing to absorption and stimulated emission. (a) A density plot for $kz = 0$ and $\Omega t = 3\pi/2$ as functions of $kx$ and $ky$. (b) Plots for $kx = 50$ and $kz = 0$ as a function of $ky$ for various values of $\Omega t$ (dotted (black): $\Omega t = 0$ and $\pi$; thick broken (red): $\Omega t = \pi/4$ and $3\pi/4$; thick solid (yellow): $\Omega t = \pi/2$; thin broken (green): $\Omega t = 5\pi/4$ and $7\pi/4$; thin solid (blue): $\Omega t = 3\pi/2$). Density plots for $kx = 50$ are given as functions of $ky$ and $kz$ for (c) $\Omega t = 3\pi/2$ and (d) $\Omega t = \pi/2$. Here, $\eta_i(\boldsymbol{r},t)$ is normalized by the constant $0.01A$.

Figure 2. Spatial distribution of $\eta_r(\boldsymbol{r},t)$ – component of interference term $\eta(\boldsymbol{r},t)$ contributing to dispersion, where (a, c) $\Delta > 0$ and (b, d) $\Delta < 0$. Here, (a) and (b) show plots for $kx = 50$ and $kz = 0$ as a function of $ky$ for various values of $\Omega t$ (dotted (black): $\Omega t = 0$; thick broken (red): $\Omega t = \pi/4$ and $7\pi/4$; thick solid (yellow): $\Omega t = \pi/2$ and $3\pi/2$; thin broken (green): $\Omega t = 3\pi/4$ and $5\pi/4$; thin solid (blue): $\Omega t = \pi$). (c) and (d) are density plots for $kx = 50$ and $\Omega t = \pi$ as functions of $ky$ and $kz$. Here, $\eta_r(\boldsymbol{r},t)$ is normalized by the constant $0.01A|\Delta|/\Omega$.

Figure 3. Density plots of interference term $\eta(\boldsymbol{r},t)$ with $kx = 50$ for $\Delta = \Omega_0$ as functions of $ky$ and $kz$ ($-50 < ky < 50$, $-50 < kz < 50$), where (a) $\Omega t = 0$, (b) $\Omega t = \pi/4$, (c) $\Omega t = \pi/2$, (d) $\Omega t = 3\pi/4$, (e) $\Omega t = \pi$, (f) $\Omega t = 5\pi/4$, (g) $\Omega t = 3\pi/2$ and (h) $\Omega t = 7\pi/4$. Here, $\eta(\boldsymbol{r},t)$ is normalized by the constant $0.01A$.

Figure 4. Schematic pattern of light waves under coherent interaction with an atom.

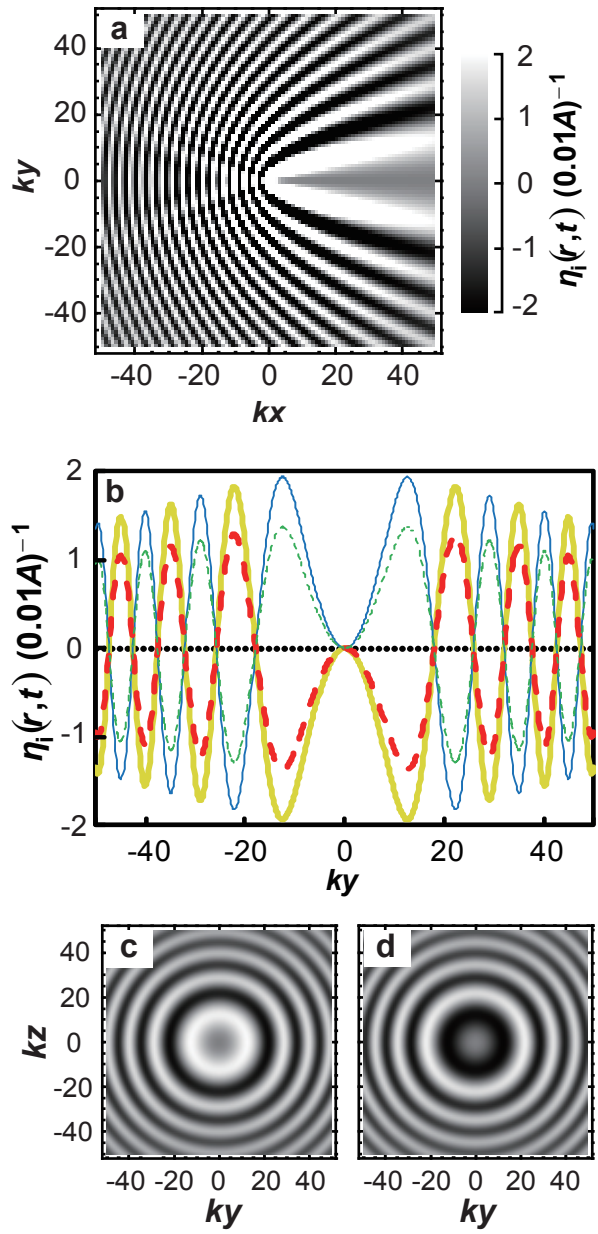

Fig. 1. A. Takamizawa et al.

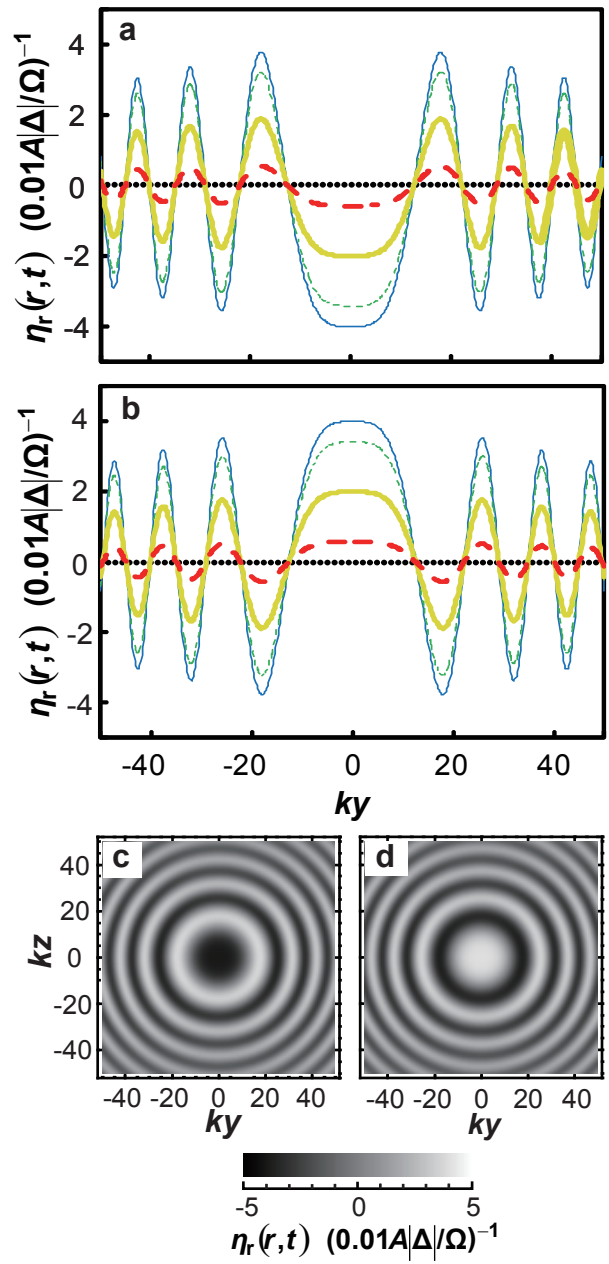

Fig. 2. A. Takamizawa et al.

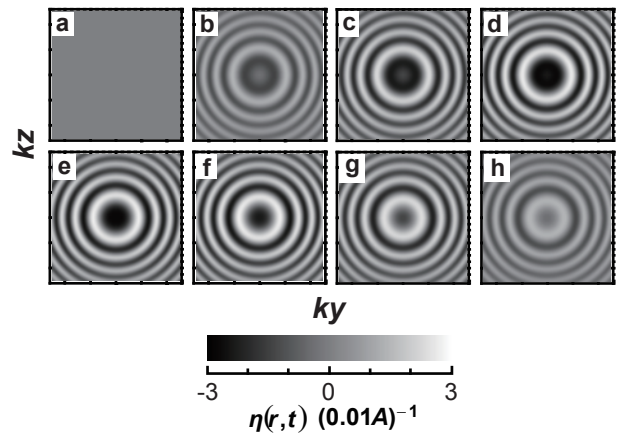

Fig. 3. A. Takamizawa et al.

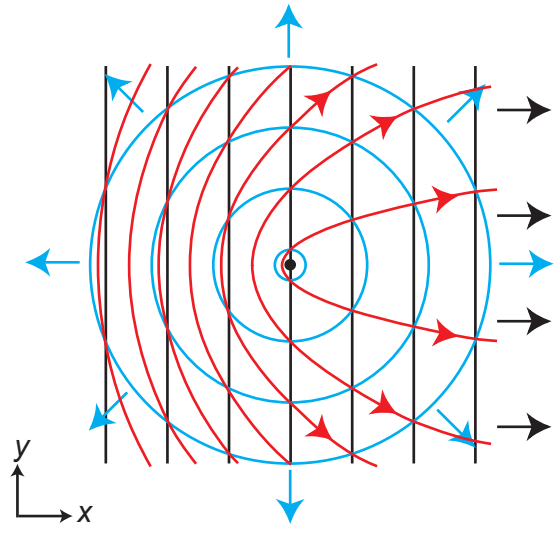

Fig. 4 A. Takamizawa et al.